\documentclass[a4paper,11pt]{article}
\pdfoutput=1 

\usepackage{jheppub} 

\usepackage[T1]{fontenc} 

\usepackage[usenames,dvipsnames]{xcolor}

\usepackage{dcolumn}   
\usepackage{amsmath}
\usepackage{mathtools}
\usepackage[T1]{fontenc}
\usepackage{amssymb}
\usepackage{amsfonts}
\usepackage{amsthm}
\usepackage{bm}
\usepackage{mathrsfs}
\usepackage{slashed}
\usepackage[all]{xy}
\usepackage{tensor}
\usepackage{stmaryrd}
\usepackage{slashed}

\usepackage[usenames,dvipsnames]{xcolor}

\newcommand{\vep}{\varepsilon} 

\newcommand{\ep}{\epsilon}
\newcommand{\Bif}{\mathcal{B}}
\newcommand{\Hor}{\mathcal{H}}
\newcommand{\Sph}{\mathcal{S}}

\newcommand{\beq}{\begin{equation}}
\newcommand{\eeq}{\end{equation}}

\title{\boldmath The integrability of Virasoro charges for axisymmetric Killing horizons}

\author{Lin-Qing Chen}
\affiliation{ Centre for Quantum Information and Communication, Ecole polytechnique de Bruxelles,
	CP 165, \\Universit\'e Libre de Bruxelles, 1050 Brussels, Belgium}

\abstract{Through the analysis of null symplectic structure, we derive the condition for integrable  Virasoro generators on the covariant phase space of  axisymmetric Killing horizons. A weak boundary condition  selects a special relationship between the two temperatures for the putative CFT. When the integrability  is satisfied for both future and past horizons, the two central charges are equal. At the end we discuss the physical implications.}

\emailAdd{linqing.nehc@gmail.com}

\begin{document} 
\maketitle
\section{Introduction}
What are the degrees of freedom responsible for the horizon entropy? How can we define their phase space, and eventually, their Hilbert space?  The presence of horizon as a boundary relative to the outside observers promotes certain gauge degrees of freedom to be physical degrees of freedom on the phase space \cite{Carlip:1994gy, Balachandran:1994up, Teitelboim:1995fr, Donnay:2015abr, Hawking:2016msc,  Hawking:2016sgy}.  It manifests as the existence of nontrivial quasilocal Hamiltonian $H_\xi$  generating a flow for any phase space functions $F$ 
\beq
\{ H_\xi , F\} = \delta_{\xi} F,
\eeq
associated with the large diffeomorphisms given by the vector field $\xi$.   Intuitively, such type of Hamiltonian flow transforms through different states on the phase space paramatrized by the boundary degrees of freedom, which would be pure gauge if the partition of a boundary was lifted \cite{Balachandran:1994up, Donnelly:2016auv, Speranza:2017gxd, Barnich:2007bf, Compere:2018aar}. 
The preservation of a different boundary structure on the horizon corresponds to a different symmetry  \cite{Barnich:2007bf, Compere:2018aar,  Afshar:2016wfy, Donnay:2016ejv}.  For 2d conformal algebra and  related symmetries, when there exists a corresponding Cardy-like formula \cite{Cardy:1986ie, Bagchi:2012xr, Detournay:2012pc}, although extra assumptions such as modular invariance  are needed, in many cases it has been shown that the Cardy entropy remarkably coincides with the Beikenstein-Hawking entropy \cite{Strominger:1997eq, Carlip:1998wz, Carlip:1999cy, Carlip:2017xne, Carlip:2019dbu, Haco:2018ske, Aggarwal:2019iay, Haco:2019ggi, Chen:2020nyh}, which suggests a CFT intepretation of the horizon microstate.

 This paper aims at strengthening our previous work by deriving a substantial assumption made in the paper \cite{Chen:2020nyh}.  In our previous work, by generalizing the construction of \cite{Haco:2018ske, Castro:2010fd} to generic axisymmetric Killing horizons, we have shown that the central charges from the 2d conformal algebra of near-horizon phase space are proportional to the area of bifurcation surface: $c_R=c_L = 3 A /2\pi G(\alpha+\beta)$ \cite{Chen:2020nyh}.  However,  to show that the Virasoro generators could actually be integrable Hamiltonians on the covariant phase space was left for future study.  A rigurous justification of integrability  is the quest the present paper aims to address.

It was initially expected that it is necessary to evoke the Wald-Zoupas counterterm prescription  in order to make the Virasoro charges well-defined \cite{Wald:1999wa, Haco:2018ske, Haco:2019ggi,  Perry:2020ndy}.  A comprehensive analysis was carried out in  \cite{Chandrasekaran:2020wwn}  by using the Barnich-Troessaert bracket for nonintegrable charges due to the gravitational flux \cite{Barnich:2011mi}. The present article focuses on the situation in which  the condition for  integrable charges does exist  without the need for a counterterm.  The question is nonconventional, as the large diffeomorphisms generated by the conformal algebra lead to nontrivial variation on both of the surface gravity and the horizon generator. By analyzing the null symplectic structure, finally we arrive at a weak boundary condition which provides a special relationship between two CFT temperatures, as well as selects a particular family of the $AdS_3$ folia in the near horizon region. It automatically leads to $c_L =c_R$ when the integrability condition is satisfied.




\section{Charges, ambiguities, integrability and the central extension }
In this section, we will review the covariant phase space formalism with special emphasis on the integrability, central extension and the ambiguities \cite{Iyer:1994ys, Iyer:1995kg, Wald:1999wa, Compere:2018aar, Barnich:2001jy, Harlow:2019yfa, Ruzziconi:2019pzd, Shi:2020csw}.

Starting from a classical action defined on the spacetime $M$\footnote{In this article, we use bold letters to denote densities.}: $S = \int_M \bold{L} + \int_{\partial{M}} \bm{l}_{b}$, the variation of the action tells us about both the equation of motion and the symplectic structure of the phase space:
\beq
\delta S = \int_M \bold{E}(g) \delta g + \int_{\partial{M}} (\bm{\theta}+ \delta \bm{l}_{b}).
\label{action}
\eeq
In our case  $\bold{E}(g) $ is the Einstein equation. 
$\bm{\theta}(g, \delta g)  $ is  the symplectic potential density, which is a spacetime $d-1$-form, field space 1-form.  Here, all the tensor indices are omitted. The choice of the boundary term of the action $\bm{l}_{b}$ is crucial for a well-defined variational principle.  When $\partial M$ is a time-like boundary (for a comprehensive analysis see \cite{Harlow:2019yfa}), $\bm{l}_{b}$ shall be the Gibbons-Hawking boundary term.  When $\partial M$ is a null boundary, there are various proposals in the literature.  As a field space exact form, such a boundary term corresponds to a change of polarization for certain pairs of canonical variables: $p\wedge \delta q \rightarrow -\delta  p \wedge q$

The presymplectic 2-form is an integration on a Cauchy surface $\Sigma$:
\beq
\Omega(g, \delta_1 g,  \delta_2 g ) = \int_{\Sigma}\delta_1 \bm{\theta} (g, \delta_2 g ) - \delta_2 \bm{\theta} (g, \delta_1 g ).
\eeq
We are studying the Hamiltonians (charges) which generate nontrivial transformations $\delta_{\zeta} F(g_{ab})$ of phase space function $F(g_{ab})$  induced by diffeomorphisms $\delta_{\zeta} g_{ab} := \pounds_\zeta  g_{ab}$.  
The trivial gauge transformation corresponds to the degenerate direction of the symplectic form: $\Omega(g, \delta g,  \pounds_\zeta  g_{ab})=0 $, while the nontrivial one will have corresponding quasi-local charges:
\beq
\slashed{\delta} H_{\zeta} = \Omega (\delta g_{ab}, \delta_\zeta g_{ab})= \int_{\partial{\Sigma}} \delta \bm{Q}_{\zeta} - \iota_\zeta \bm{\theta} (g_{ab}, \delta g_{ab}   ) ,  
\label{slashedH}
\eeq
where $Q_{\zeta}$ is the \textit{Noether potential} (d-2) form, or  the \textit{Komar charge} in gravity context: 
\beq
d \bm{Q}_{\zeta} :=\bm{j}_\zeta =  \bm{\theta} (g_{ab}, \delta_\zeta g_{ab} )  - \iota_\zeta \bm{L} ,
\eeq
up to terms that are proportional to the equation of motion. For local and covariant theory, $d\bm{j}_\zeta=0$ onshell. 

In the consideration of adding a Lagrangian boundary term $\bm{l}_{b}$,  the  symplectic potential becomes $\bm{\theta} + \delta \bm{l}_{b}$, the Noether current (d-1)-form would be modified as:
\beq
\tilde{\bm{j}}_{\zeta} =  \bm{j}_{\zeta}   +d \iota_\zeta \bm{l}_{b} + \delta_\zeta \bm{l}_{b} - \pounds_\zeta \bm{l}_{b}. 
\label{current}
\eeq
Hence the Komar charge $\bm{Q}_{\zeta}$ would be shifted by $\iota_\zeta \bm{l}_b$.  If the Lagrangian boundary term $\bm{l}_b$ is noncovariant, $d\bm{j}_\zeta\neq 0$, which is the case in our study.

The symplectic potential density $\bm{\theta}$ is only defined by the action up to a total derivative, which is part of the JKM ambiguities \cite{Jacobson:1993vj}.  Its physical significance was recently studied in \cite{Freidel:2020xyx}. Consider a shift of the symplectic potential density by a spacetime exact form:
\beq
\bm{\theta} \rightarrow\bm{\theta} + d \bm{b }(g, \delta g). 
\label{thetashift}
\eeq
Such a transformation would leave the equation of motion invariant, but modify the variation of the Hamiltonian  (\ref{slashedH}) by 
\beq
\slashed{\delta}H_\zeta \rightarrow \slashed{\delta}H_\zeta  + \int_{\partial{\Sigma}} \delta  \underbrace{ \bm{b }(g, \pounds_{\zeta} g) }_{\text{shift}  \ \bm{Q}_\zeta}- \pounds_\zeta  \bm{b }(g, \delta g).
\eeq
Here the slashed variation $\slashed{\delta} H_{\xi}$ emphasizes that  the Hamiltonian can only be well-defined if it is an exact form on the field space, i.e. $\delta \slashed \delta H=0$. The reason is that to determine the Hamiltonian $H_{\xi}$ from the expression of its variation (\ref{slashedH}), it needs to be independent of the path of integration between any reference spacetime metric $g_o$ to $g$ in the phase space:
\beq
H_\zeta (g_o, g)= \int_{g_o} ^ g  \delta H_{\zeta} + N_{\zeta}(g_o),
\label{path}
\eeq
which is only true if $\slashed{\delta} H_{\zeta}$ is a field-space exact form -- then we say that the Hamiltonians (charges) are \textit{integrable}.  In the above equation, $N_{\zeta}(g_o)$ is an integration constant and it may be influenced by both $\delta \bm{l}_b$ and $d\bm{b}(g,\delta g)$. We will evaluate such modification at the  later part of the paper.

The most direct approach to check the integrability is by the following necessary condition:
\begin{equation}
\delta ( \slashed{\delta}H_\zeta )= -\delta  \int_{\partial \Sigma} \iota_\zeta \bm{\theta} (g,\delta g)=
- \int_{\partial \Sigma'} \iota_\zeta \bm{\omega}(g,\delta_1 g,\delta_2 g) \overset{!}{=}0.
\end{equation}
However, sometimes this expression can be cumbersome as the symplectic form has double the number of the field variation than the symplectic potential. A more controlled way is by analyzing the obstruction terms  within the (\ref{slashedH}) directly. The subtleties of identifying the integrable part of the charges have been discussed in, for example \cite{Compere:2018ylh, Compere:2020lrt, Fiorucci:2020xto}.


\textit{If the Hamiltonian is integrable}, then given a set of vector fields ${\zeta_m}$ (we assume that they are field-independent in the scope of this work, i.e. $\delta \zeta_m =0$), by the charge representation theorem \cite{Barnich:2007bf, Compere:2018aar} , the algebra formed by the Hamiltonians  under Poisson brackets is isomorphic to the Lie algebra of diffeomorphisms  defined by the vector fields up to a central extension 
\begin{equation}
\{H_{\zeta_m}, H_{\zeta_n}\} = H_{[\zeta_m, \zeta_n]} + \mathcal{K}_{m,n},
\end{equation}
where
$ \mathcal{K}_{m,n} $ is given by
\begin{equation}
\mathcal{K}_{m,n}(g)  =	\int_{\partial{\Sigma}}  \iota_{\zeta_m} \bm{\theta}(g,\pounds_{\zeta_n} g)  -\iota_{\zeta_n} \bm{\theta}(g,\pounds_{\zeta_m} g)  + \iota_{\zeta_n} \iota_{\zeta_m} \bm{L}.
\label{K}
\end{equation}
The central extension measures the \textit{difference}, or in another word, an \textit{anomaly} on representing the Lie algebra of the vector fields by the Poisson algebra on the phase space. 
It is a Casmir on the phase space -- does not generate any flow. It forms the two-cocycle on the Lie algebra of diffeomorphisms
\beq
\mathcal{K}_{[m,n],k} + \mathcal{K}_{[k,m],n} +\mathcal{K}_{[n,k],m} =0.
\eeq

\section{The null symplectic structure }
\subsection{Review of the symplectic structure on the null hypersurface}
The symplectic structure of gravity on null hypersurface has a very elegant and simple form in terms of geometrical quantities \cite{Ashtekar:1981bq, Hopfmuller:2016scf, Parattu:2015gga, Lehner:2016vdi, Hopfmuller:2018fni, Chandrasekaran:2018aop, Hayward:1993my, Poisson:2009pwt}.  This section serves as a brief review \cite{Hopfmuller:2016scf, Hopfmuller:2018fni}. 
Let us start from the symplectic potential density (d-1)-form from the Einstein-Hilbert action with cosmological constant:
\beq
\bm{L} =\frac{1}{16\pi G}  (R -2 \Lambda ) \epsilon
\eeq
\beq
\bm{\theta} (g_{ab}, \delta g_{ab}) = \frac{1}{16\pi G} \ep^a\left(\nabla^b\delta g_{ab}-
g^{bc} \nabla_a \delta g_{bc} \right),
\eeq
where  $\epsilon$ denotes the spacetime volume form, and $\epsilon_a = \iota_a \epsilon$ is the directed (d-1)-form. Its pullback onto the null hypersurface $\Hor$ with boundary $\partial \Hor$ can be written as an integration on $\Hor$ and a corner piece: $\Theta_\Hor + \Theta_{\partial \Hor} $.
The first part $ \Theta_\Hor$ only depends on the intrinsic and extrinsic geometry of the null hypersurface $\Hor$, and can be writen in the form of canonical pairs $``p\delta q"$:
\begin{equation}
 \Theta_\Hor= \frac{1}{16 \pi G} \int_\Hor \left(  \underbrace{ \sigma^{ab}\delta \gamma_{ab}}_{\text{Spin-2}} - 2\underbrace{\omega_a \delta \ell^a}_{\text{Spin-1}} - \underbrace{(\kappa + \frac{d-3}{d-2}\theta)\delta  \tilde{q} }_{\text{Spin-0}} \right) \epsilon_\Hor + \frac{1}{8 \pi G} \int_\Hor \delta(\kappa \epsilon_\Hor),
 \label{bulkfull}
\end{equation}
in which $\epsilon_\Hor$ denotes the volume form of the null hypersurface.  In the spin-2 term, the configuration varible is given by the  conformal metric $\gamma_{ab}$ of spacial cross-section and the conjugate moment can be identified as \textit{shear} $\sigma^{ab}$. For the spin-1 part, the configuration variable is the generator $\ell^a$ of the null hypersurface, while the conjugate momentum is the \textit{twist} $\omega_a$ 1-form. For the spin-0,   $\tilde{q}$ is the trace of the  induced metric on the co-dimention-2 surface and the conjugate momentum is a combination of the expansion $\theta$ and the surface gravity $\kappa$.

The last term $\delta (\kappa \epsilon_\Hor)$ is a total variation (exact form) on the field space. Including it into the  integration on $\Hor$  changes the polarization on the spin-0 piece of the symplectic potential. Such an exact form can be modified by adding a boundary term $\bm{l}_{b}$ to the Lagrangian and does not influence the equation of motion.

The corner piece  $\Theta_{\partial \Hor} $ could be sorted into different forms depending on the geometrical feature one wants to study. One convenient expression is explicitly given by
\begin{equation}
\Theta_{\partial \Hor} =\frac{1}{16 \pi G} \int_{\partial{\Hor}}  \left((1+ \frac{1}{2}h)  \delta \tilde{q} \ell^a    + (1+ h) \delta\ell^a  \right)\iota_a \epsilon_\Hor  -\frac{1}{16 \pi G}\int_{\partial{\Hor}}  \delta \left( h \ell^a \iota_a \epsilon_\Hor \right),
\label{boundary}
\end{equation}
where the factor $h$ is defined by the normal volume element:
\beq
e^h := \sqrt{ |g|}/\sqrt{\tilde{q}}.
\label{red}
\eeq
The factor $h$ can be identified as a \textit{red shift factor} of light rays  parallel to the null hypersurface measured by a geodesic observer crossing the hypersurface. Please see the section VA in \cite{Hopfmuller:2016scf} on more physical discussion regarding this factor. 


\subsection{Analysis in the context of axisymmetric Killing horizons} 
\label{sectionkilling}
To have a general understanding of how geometrical structures change under the metric variation, we take a  perspective that  coordinates are “fields" to define foliations on the spacetime manifold \cite{Hopfmuller:2018fni}.  Under active diffomorphisms, the label of foliation does not change.  The location of the horizon is specified by  $X=0$ and hence fixed. Constant $X$  foliations are either timelike or null. Constant $T$ foliations label spacelike hypersurfaces and  can be viewed as a “clock” along the horizon. To be well-defined foliations, they are all linearly independent and $dT \neq 0, dX \neq 0$.  There is no spacetime metric information involved in the foliation fields $X, T, \sigma_i$, hence $\delta T, \delta X, \delta \sigma_i =0$. 

We study  the spacetime metric  near a stationary, bifurcate, axisymmetric Killing horizon in $d\geq 3$. The horizon generators $\chi^a = t^a + \Omega_\Hor \psi^a$ are the linear combination of the timelike killing vector $t^a$ and the rotational Killing vector $\psi^a$. In the Rindler form of coordinates, the metric can be expressed as 
\beq
ds^2 = -\kappa^2 x^2 dt^2 +  dx^2 + \psi^2 d\phi^2 
+ q_{AB} d\theta^A d\theta^B - 2 \kappa x^2 dt \big( N_\phi d\phi  
+ N_A d\theta^A \big) + ..., \label{Rind}
\eeq
where $x$ is the proper distance to the bifurcation surface at the leading order, $\phi$ is the comoving angular coordinate corresponding to the rotational killing vector and $\theta^A$ lables all the rest of angular coordinates.  Note that (\ref{Rind}) omitted certain irrelevant terms at the order of $\mathcal{O}(x^2)$, for the full details see appendix of  \cite{Chen:2020nyh}.   In the Kruskal coordinates $(U,V)$:
\beq
U = x e^{\kappa t}, \ \ V=  x e^{-\kappa t},
\eeq
the  metric (\ref{Rind}) has the following form
\beq
ds^2 = dU dV -  U dV (N_\phi d\phi+ N_A d\theta^A ) +   V dU (N_\phi d\phi+ N_A d\theta^A ) + \psi^2 d\phi^2 
+ q_{AB} d\theta^A d\theta^B + ...
\eeq
The past horizon $\Hor^-$ and future horizon $\Hor^+$ are located at $U, V=0$ respectively, with a bifurcation surface $\Bif$ connecting $\Hor^+_-$ and $\Hor^-_+ $. 

In the Rindler form, although the redshift factor $h=\ln \left(\sqrt{ |g|}/\sqrt{\tilde{q}} \right)$ is infinite on the horizon, the coordinates $(t,x)$ can still play the role of a good foliation function as $dt, dx\neq 0$. Given any spacelike cross-section  $\Sph$ of the horizon, the normal derivatives covariant under diffeomorphisms of $\Sph $ are defined as
\beq
D_i := \partial_i + A_i^\sigma \partial_{\sigma},  i \in\{t,x\}, \sigma \in \{\phi, \theta^A\},
\eeq
where $\sigma$ labels the coordinate indices on the codimension-2 cross-section $\Sph$. In our case,
\beq
 A_t^\phi = \kappa x^2 N_\phi /\psi^2, \ \  A_t^\theta = \kappa x^2 N_A q^{AB},\ \  A_x^\phi =A_x^\theta = 0.
\eeq
For any spatial cross-section $\Sph$ of the horizon with induced metric $\tilde{q}_{ab}$,  following the light rays, there are two null vectors normal to $\Sph$: $\ell^a \tilde{q}_{ab} = \bar{\ell}^a \tilde{q}_{ab}=0$. One is intrinsic to the horizon, the other  is transverse to the horizon and we use  $\bar{\ell}$ to label it. For the future horizon $\Hor^+$, they  can be defined as
\beq
\ell^a \partial_a = \frac{1}{2} (D_t + \kappa x D_x  ) \overset{H}{=} \frac{1}{2} \kappa U \partial_U,\ \ \bar{\ell}_a dx^a =  dt + \frac{1}{ \kappa x}  dx= (\kappa U)^{-1} dU.
\label{null1}
\eeq
For the past horizon  $\Hor^-$,  we use $n^a$ to label the intrinsic null vector  and the transverse one is labeled by $\bar{n}^a$:
\beq
n^a \partial_a = \frac{1}{2} (\kappa x D_x -D_t  ) \overset{H}{=} \frac{1}{2} \kappa V \partial_V,\ \ \bar{n}_a dx^a =- dt +   \frac{1}{ \kappa x}  dx = (\kappa V)^{-1} dV.
\label{null2}
\eeq
The induced metric $\tilde{q}_{ab}$ on $\Sph$ can be expressed in terms of its two null normals and the spacetime metric $ g_{ab} $: 
\begin{equation}
\tilde{q}_{ab} = g_{ab} -  \ell_a  \bar{\ell}_b -   \bar{\ell}_a  \ell_b = g_{ab} -  n_a  \bar{n}_b -   \bar{n}_a  n_b.
\end{equation}
One can readily check  that $\ell^a q_{ab} \sim  \mathcal{O}(x^4)$.  The null vector and covector  satisfy a convention of normalization $\ell^a \bar{\ell}_a =n^a \bar{n}_a =1$. Note that, in general such normalization fixes the full normals up to a rescaling of an arbitrary function  $\epsilon$: $e^{\epsilon} \ell^a , e^{-\epsilon} \bar{\ell}_a$, which is called \textit{boost gauge} in literature.   The form of definition (\ref{null1}, \ref{null2}) is to ensure that each term in the symplectic potential is boost gauge invariant respectively \cite{Hopfmuller:2016scf}.  Such choice of null vectors defines a notion of a "clock" along the null ray intrinsic to the horizon: $\ell [T] = \text{constant}$, which is the only fixed structure in the analysis\footnote{This is different from the related work \cite{Chandrasekaran:2020wwn} as  $\delta \ell_a =0$ is not imposed as the fixed structure. There is also no constraint on $\bar{\ell}^a \ell^b \delta g_{ab}$.  As a result, the field space variation of surface gravity $\delta_\zeta \kappa$ will be different, as we will see in the next section.}, hence,
\beq
\bar{\ell}_a \delta \ell^a  = 0
\eeq
and $\delta \ell^a $ is parallel to the horizon cross-section. 
 As it was initially emphasized in \cite{Hopfmuller:2018fni}, fixing the time foliation introduces a sense of noncovariance to the phase space as we will see in detail later. At this point, the only requirement on the metric variation is to preserve the nullness of the horizon: $\ell^a \ell^b \delta g_{ab} \overset{\Hor^+}{=} 0, \delta{\ell^a} \ell_a \overset{\Hor^+}{=} 0$, the same relation holds for $n^a$ on the past horizon.

Now let us look at the symplectic potential, with the possibility of adding a  boundary action $\bm{l}_b$ but with an undetermined coefficient $\eta$:
\beq
S = \frac{1}{8\pi G}   \int_M  (R/2 - \Lambda ) \epsilon + \frac{\eta}{8\pi G} \int_\Hor \bm{l}_{b}.
\eeq
For killing horizons, the vanishing of expansion $\theta$ and shear $\sigma^{ab}$  simplifies the symplectic structure \footnote{However one needs to be careful  as the field space variation of the expansion $\theta$ and shear $\sigma^{ab}$ might not vanish. 
It depends on whether the physical situation we are considering would preserve the null hypersurface to be a Killing horizon.}.  The bulk piece of the symplectic potential (\ref{bulkfull})  becomes
\begin{equation}
\Theta_\mathcal{H}  = \frac{1}{8 \pi G}\int_\mathcal{H} \left( -  \omega_a \delta l^a + \delta  \kappa  \right) \epsilon_\mathcal{H} ,
\end{equation}
in which $\omega_a$ is the  \textit{twist} and it  describes how a surface twists inside the horizon if we let it move along the integral curve of the normal vector $\ell$:
\beq
\omega_a : = -\tilde{q}_a^{\ b} \ell^c \nabla_c \bar{\ell}_b .
\eeq
Geometrically $\tilde{q}_a^{\ b}$ is the projector onto  $\Sph$. Evaluated on the horizon, the twist $\omega_a dx^a$ has the simple form:
\beq
\omega_a dx^a \overset{\Hor^+}{=} N_\phi d\phi + N_A d\theta^A.
\eeq
The twist 1-form on the past horizon $\omega_a : = -\tilde{q}_a^{\ b} n^c \nabla_c \bar{n}_b$ has the equivalent evaluation, with an extra minus sign. 

We consider the boundary Lagrangian in the following form:
\beq
\bm{l}_{b} = \kappa \epsilon_\mathcal{H}.
\eeq
 Such type of  boundary term for null hypersurface has been studied in \cite{Parattu:2015gga, Lehner:2016vdi,Hopfmuller:2016scf,  Chandrasekaran:2020wwn}.   With the boundary term, the symplectic potential will have a total shift by $\Theta_\Hor + \lambda \delta  (\kappa \epsilon_\mathcal{H} )$.    As part of the symplectic potential ambiguity, another nontrivial exact form that one could consider   is $\delta(\omega_a l^a \epsilon_\Hor)$ . However, as $\omega_a l^a \overset{\Hor}{=} 0$, it does not contribute to the boundary action.    
 

The variation of horizon volume form  can be evaluated explicitly,
\beq
\delta  \epsilon_\mathcal{H} = \delta \sqrt{|\tilde{q}|} dt\wedge d^{(d-2)} \sigma= \frac{1}{2} \delta \tilde{q} \epsilon_\mathcal{H}=\frac{1}{2} \tilde{q}^{ab} \delta \tilde{q}_{ab} \epsilon_\mathcal{H}.
\eeq
Recall that $\tilde{q}_{ab} $ is the induced metric on $\Sph$ and $\tilde{q}$ represents its trace, hence the variation of the boundary Lagrangian gives
\beq
\delta \bm{l}_{b} = \delta   (\kappa \epsilon_\mathcal{H}) = (\delta \kappa) \epsilon_\mathcal{H}+  \frac{\kappa}{2} \tilde{q}^{ab} \delta \tilde{q}_{ab} \epsilon_\mathcal{H},
\eeq
which can be viewed as a mixed polarization between considering the trace of the induced metric $\delta \tilde{q} =\tilde{q}^{ab} \delta \tilde{q}_{ab} $ as the configuration variable while $\kappa$ as its conjugate momenta (\ref{bulkfull}), v.s. conversely considering $\kappa$ as the configuration variable on the phase space.

Now let us simplify the corner term (\ref{boundary}).
As $\delta \ell^a$ is parallel to $\Sph$, the following integration vanishes: $\int_\Sph \delta \ell^a \iota_a \epsilon_\Hor = 0$. 
Furthermore as $2 \delta(\iota_a \epsilon_H) =  \delta \tilde{q} (\iota_a \epsilon_H)$, the corner term of the simplectic potential becomes
\begin{equation}
\Theta_{\Sph}  = \frac{1}{16 \pi G} \int_{\Sph} (\delta \tilde{q} - \delta h) \iota_\ell \epsilon_H.
\label{bdry1}
\end{equation}
Finally, the expressions of the two Noether charges corresponding to the stationary killing vector $t^a$ and the rotational Killing vector $\psi^a$ are the area $A$  and the angular momentum $J_H$ \cite{Chen:2020nyh}: 
\beq
A = 2\pi \int d\theta^A \sqrt {|q_{AB}|}  |\psi|, \ \ 
J_H=  \frac{1}{4G} \int d\theta^A \sqrt{ |q_{AB}|}|\psi| N_\phi .
\label{J}
\eeq
To evaluate them we used the area form of the cross-section $\Sph$:  $\epsilon_S = \sqrt{|\tilde{q}|} d^{(d-2)} \sigma=  \sqrt {|q_{AB}|} |\psi| d\phi \wedge d\theta^A$  for the near horizon metric (\ref{Rind}).


\section{The Virasoro hairs}
Inspired by the earlier work of \cite{Maldacena:1998bw, Haco:2018ske, Castro:2010fd}, for any axisymmetric Killing horizons in $d\geq 3$, the following conformal coordinate transformation \cite{Chen:2020nyh}
\begin{equation}\label{con}
\begin{split}
w^+ &= xe^{\alpha \phi + \kappa t} \\
w^- &= x e^{\beta\phi -\kappa t} \\
y &= e^{\frac{\alpha + \beta}{2} \phi} 
\end{split}
\end{equation}
was designed to bring out the  $AdS_3$ folia within the near horizon geometry. Under such transformation, the metric (\ref{Rind}) becomes
\begin{equation}
\begin{split}
	ds^2   &=  \frac{dw^+ dw^-}{y^2} 
+ \frac{4 \psi^2 }{(\alpha+\beta)^2 }\frac{dy^2}{y^2}
+q_{AB}d\theta^A d\theta^B   \\
&\;	-\frac{2 dy}{(\alpha+\beta)y^3} \Big((\beta+ N_\phi  )   w^-dw^+ + (\alpha- N_\phi  )   w^+dw^-   \Big)   \\
&\; - \left(\frac{w^- dw^+}{y^2} - \frac{w^+ dw^-}{y^2}\right) \kappa N_A d\theta^A + \ldots
\end{split}
\end{equation}
up to higher order in $w^\pm$. The $(w^\pm, y)$ subspace defines a warped $AdS_3$ geometry with $\theta$ dependent radius
\beq
L_{AdS_3} = \frac{2 |\psi|} {\alpha + \beta}.
\eeq
The $\alpha, \beta$ are two arbitrary parameters, which have an interpretation of the putative CFT temperatures $(\alpha = 2 \pi T_R,\  \beta = 2 \pi T_L )$ identified from the periodicity condition and the Frolov-Thorne vacuum density matrix with respect to the Killing vectors \cite{Frolov:1989jh}.  The   asymptotic  vector fields for the $AdS_3$  folia are the Brown-Henneaux vector fields without the field-dependent term \cite{Haco:2018ske}: 
\beq\label{vf}
\zeta^a_\vep = \vep(w^+)\partial_+^a +\frac{1}{2} \vep'(w^+) y\partial_y^a
\eeq
\beq\label{vp}
\xi^a_{\bar\vep} = \bar\vep(w^-) \partial_-^a +\frac{1}{2} \bar\vep'(w^-) y\partial_y^a.
\eeq
From the periodicity of $\phi$, the basis for the mode expansion is given by $\vep_m(w^+) = \alpha (w^+)^{1+\frac{im}{\alpha}}$, 
$\bar \vep_n(w^-) = -\beta\, (w^-)^{1-\frac{in}{\beta}}$. The corresponding generators are labeled as $\zeta_m$, $ \xi_n$ respectively and their Lie algebras form two commuting copies of the Witt algebra, 
\beq
[\zeta_m , \zeta_n] = i(n-m)\zeta_{m+n}, \ \ 
[\xi_m ,\xi_n] = i(n-m)\xi_{m+n}.
\eeq
Each copy of the Witt algebra is well-defined on one horizon ($\Hor^+$ or $\Hor^-$), but they do not have regular limit on the other horizon.  It is illuminating for our analysis to decompose the vector fields in terms  of the component parallel, normal to the horizon and parallel to the spatial cross-section $\Sph$. For the set of vector field which are regular on the future horizon (\ref{vf}), $\zeta^a_m  = f_m \ell^a + \bar{f}_m \bar{\ell}^a + v^a$ and $v^a \parallel \Sph$,
\beq
\begin{split}
f_m &= \alpha  U^{\frac{im}{\alpha}} e^{i m \phi} \ \frac{\beta - i m }{\kappa (\alpha + \beta)}\\
\bar{f}_m &= - \beta   U^{\frac{im}{\alpha}} e^{i m \phi}\ \frac{ \kappa x^2(i m + \alpha )}{2 (\alpha + \beta)}\\
v^a &=  U^{\frac{im}{\alpha}} e^{i m \phi}\  \frac{i m+ \alpha}{\alpha + \beta}  \partial^a_\phi.
\end{split}
\eeq
For  the other set of vector fields $\xi_m$ (\ref{vp}) which are regular on the past horizon $\Hor^-$, similarly one could have the decomposition $\xi^a_m  = r_m n^a + \bar{r}_m \bar{n}^a + u^a$ in which
\beq
\begin{split}
	r_m &= - \beta  V^{-\frac{im}{\beta}} e^{-i m \phi} \ \frac{\alpha + i m }{\kappa (\alpha + \beta)}\\
	\bar{r}_m &=  \beta   V^{-\frac{im}{\beta}} e^{-i m \phi}\ \frac{ \kappa x^2(\beta - im )}{2 (\alpha + \beta)}\\
	u^a &=  V^{-\frac{im}{\beta}} e^{- i m \phi}\  \frac{i m - \beta}{\alpha + \beta}  \partial^a_\phi.
\end{split}
\eeq
In the above expression, all the common prefactors  come from $(w^+)^{i m/\alpha}$ or $(w^-)^{i m/\beta}$ .  One can immediately see that the transverse components of the vector fields vanish when we take the limit towards each horizon:
\beq
\begin{split}
&\Hor^+: \bar{f}_m \bar{\ell}^a 	\propto   U^{\frac{im}{\alpha}} V\partial^a_V \rightarrow 0, \text{as} \ V\rightarrow 0,  \\
&\Hor^-: \bar{r}_m \bar{n}^a 	\propto   V^{-\frac{im}{\beta}} U\partial^a_U \rightarrow 0, \text{as} \  U\rightarrow 0.
\end{split}
\eeq 
On $\Hor^+$, the vector field $\zeta^a_m$ generates a $U$-dependent translation $f_m l^a$ plus a $U$-dependent rotation $v^a$. However, when we approach the bifurcation surface,  $U^{im}$ has singular limit as $U \rightarrow 0$. Vice versa, the other copy of the vector fields $\xi^a_n$ is regular on the past horizon and has singular limit when  $V \rightarrow 0$. As it has been observed in \cite{Haco:2018ske}, the nontrivial central extension comes from the $1/x$ pole  (or $1/w^{\pm}$ pole)  in the large diffeomorphisms, more precisely, 
\beq
 m (x e^{\kappa t})^{\frac{im}{\alpha}} e^{i m \phi} \psi^2 \frac{m- i \alpha}{x \alpha (\alpha+\beta)}\subset\pounds_{\zeta_m} g_{ab} .
 \label{singular}
\eeq

\subsection{The field variations}
This section provides the detailed results regarding the field variation of relevant geometrical quantities, which have appeared in the symplectic structure.
We use $\delta_\zeta$ to refer to a field variation induced by the diffeomorphisms generated by the  vector field $\zeta^a$, i.e. for any phase space function $T(g_{ab})$:
\beq
\delta_\zeta T(g_{ab}) = \frac{\partial T(g_{ab})}{\partial{g_{ab}}} \pounds_\zeta g_{ab}.
\eeq
It is convenient to introduce the following quantity \cite{Hopfmuller:2018fni},
\beq
\Delta_\zeta : = \delta_\zeta - \pounds_\zeta
\label{anomaly}
\eeq
which measures the difference between the variation of the phase space function generated by $\xi$ with the corresponding Lie derivative.  A convenient property is that it  commutes with the  covariant derivatives  \cite{Hopfmuller:2018fni}
\beq
\Delta_\zeta \nabla_a T(g_{ab}) =  \nabla_a  \Delta_\zeta T(g_{ab}),
\label{property}
\eeq
 which can be proven by directly checking the variation of the Christoffel symbols.

As we have discussed in the last section, the variations of the intrinsic null vector $\ell^a$ and $n^a$ induced by diffeomorphisms are parallel to the horizon cross-section $\Sph$:
\beq
\delta_{\zeta} \ell^a = - \tilde{q}^{ab} l^c \pounds_{\zeta} g_{bc}  = (\delta_{\zeta} A^\phi_t) \partial^a_\phi .
\eeq
Hence we can obtain the field space variations of $\ell^a, n^a$ generated by the vector field (\ref{vf}, \ref{vp})
\beq
\mathcal{H}^+: \ \ \ \delta_{\zeta_m} \ell^a  =  U^{\frac{im}{\alpha}} e^{i m \phi}\  \frac{m \kappa (m-i \alpha)}{\alpha (\alpha + \beta)} \partial^a_\phi  
\label{dl}
\eeq
\beq
\mathcal{H}^-: \ \ \ \delta_{\xi_m} n^a  = - V^{\frac{-im}{\beta}} e^{-i m \phi}\  \frac{m \kappa (m+i \beta)}{\beta (\alpha + \beta)} \partial^a_\phi.
\eeq
It is interesting to point out that (\ref{dl}) is exactly proportional to the coefficient of the pole in the singular diffeomorphism (\ref{singular}). For the intrinsic null vectors, the differences $\Delta_{\zeta} \ell^a $  between the field space variation $\delta_{\zeta} \ell^a$ and the Lie derivatives $\pounds_{\zeta} \ell^a$ are parallel to $\ell^a$:
\beq
\mathcal{H}^+: \ \ \ \Delta_{\zeta_m} \ell^a  =  U^{\frac{im}{\alpha}} e^{i m \phi}\  \frac{m \left(m +  i  \beta\right)} {\alpha + \beta} \ell^a
\eeq
\beq
\mathcal{H}^-: \ \ \ \Delta_{\xi_m} n^a   =-  V^{\frac{-im}{\beta}} e^{-i m \phi}\  \frac{m \left(m - i  \alpha\right)} {\alpha + \beta} n^a.
\eeq
Those quantities will facilitate the evaluation of the variation of surface gravity $\delta_\zeta \kappa$, which can be obtained by directly varying its definition $\ell^a\nabla_a \ell^b= \kappa \ell^b$ while at the same time using the property (\ref{property}):
\beq
\delta_{\zeta} \kappa = \left( (\delta_{\zeta}  \ell^a) \nabla_a \ell^b  + l^a \nabla_a (\Delta_\zeta \ell^b)  + \ell^a  \pounds_\zeta (\nabla_a \ell^b) -\kappa \delta_{\zeta} \ell^b \right) \bar{\ell}_b.
\eeq
It looks complicated at the first glance, but is actually quite simple to evaluate as we have  all the quantities in each term.  The first and third term cancels each other and we arrived at:
\beq
\delta_{\zeta_m} \kappa = l^a \nabla_a (\Delta_{\zeta_m} \ell^b)   \bar{\ell}_b \overset{\ \mathcal{H}^+}{=} m U^{\frac{im}{\alpha}} e^{i m \phi} \frac{(i m + \alpha) (m+ i\beta)}{\alpha (\alpha + \beta)},
\label{dkappa}
\eeq
\beq
\delta_{\xi_m} \kappa = n^a \nabla_a (\Delta_{\xi_m} n^b) \bar{n}_b \overset{\ \mathcal{H}^-}{=}  - m  V^{\frac{-im}{\beta}} e^{-i m \phi} \frac{(i m + \alpha) (m+ i\beta)}{\beta (\alpha + \beta)}.
\eeq
One can see immediately that for $m=0$, as the vector field is a combination of Killing vectors, the surface gravity is invariant. Interestingly, due to the fixing of different structure on the horizon, we get a different result regarding $\delta_{\zeta} \kappa $ compared to \cite{Chandrasekaran:2020wwn}  by the diffeomorphisms generated by the same vector field.

Finally, we can evaluate the field variation of the red-shift factor (\ref{red}) which appears in the corner term of the symplectic potential: 
\beq
\Hor^+: \delta_{\zeta_m}  h = U^{\frac{im}{\alpha}} e^{i m \phi}\  \frac{m \left(m -  i  \alpha\right)} {\alpha + \beta}, \ \ 
\Hor^-: \delta_{\xi_m}  h =-V^{\frac{-im}{\beta}} e^{- i m \phi}\  \frac{m \left(m + i  \beta \right)} {\alpha + \beta} .
\eeq

\subsection{The integrability condition and the central charges}
Now our goal is to identify the Hamiltonians $H_{m}, \bar{H}_n$ on the covariant phase space which generate the large diffemorphisms corresponding to $\zeta_m$ and  $\xi_n$.  We assume the Witt algebra is field-independent, i.e. they are unchanged under the variation of the metric $\delta \zeta_m = 0$.  On the future horizon, since $\zeta_m$ has a singular limit when approaching the bifurcation surface, the Hamiltonian shall be defined on the cross-section $\Sph$ of the horizon for finite $U = x e^{\kappa t}$, but \textit{not} on the bifurcation surface where $U=0$:
\beq
\slashed{\delta} H_{m} = \int_{\Sph}\delta Q_{\zeta_m} - \iota_{\zeta_m} \Theta_\Hor.
\label{slashedH2}
\eeq
With the consideration of adding a boundary term to the lagrangian, $Q_{\zeta_m}$ will be modified according to (\ref{current}).
The second term in (\ref{slashedH2})  is explicity given by
\beq
\iota_{\zeta_m} \Theta_\Hor = \frac{1}{16 \pi G}\int_{\Sph} ( -  \omega_a \delta \ell^a + \delta  \kappa + \underbrace{\eta (\delta \kappa + \frac{\kappa}{2} \tilde{q}^{ab} \delta\tilde{q}_{ab} ) }_{\text{from the boundary action}}) f_m  \epsilon_S  +\frac{1}{16 \pi G} \int_{\Sph} (\delta \tilde{q} - \delta h) \underbrace{\iota_{\zeta_m}  \iota_\ell \epsilon_\Hor}_{\text{integrated to 0}}.
\eeq
Note that the extra factor of $1/2$ comes from our normalization convention $\ell^a \partial_a dt = 1/2$.  As we can see here due to the interior product of the volume form $\iota_{\zeta_m}  \iota_\ell \epsilon_\Hor$, the corner piece vanishes when integrated on $\Sph$.  Hence the obstruction piece $\int_{\Sph}   \mathbf{O}_\eta (g, \delta g ) f_m \epsilon_\Sph$  prevents  (\ref{slashedH2}) to be an exact form,
\beq
\mathbf{O}_\eta(g, \delta g ) :=  -  \omega_a \delta \ell^a + \delta  \kappa + \eta (\delta \kappa + \frac{\kappa}{2} \tilde{q}^{ab} \delta\tilde{q}_{ab} ). 
\label{obst}
\eeq
A weak boundary condition could successfully eliminate the non-integrable part by  requiring that the obstruction term ($\ref{obst}$) integrated along the transverse dimension $\theta^A$ form an axisymmetric function in terms of $\phi$ on any cross-section of the horizon:
\beq
\partial_\phi \int  \mathbf{O}_\eta (g, \delta g ) \sqrt{|\tilde{q}|} d\theta^A  \overset{!}{=}0.
\label{bdry}
\eeq
For the metric variation generated by diffeormophisms $\delta_\zeta g = \pounds_\zeta g$, such a boundary condition will select a class of vector fields within (\ref{vf}) by the constraint on $\alpha$ and $\beta$.  As those parameters are related to the CFT temperatures by $\alpha = 2 \pi T_L, \beta = 2 \pi T_R$, it corresponds to a specific relationship between the $T_L, T_R$, area $A$ and spin $J_H$ for the horizon.
On the future horizon,  (\ref{bdry}) gives rise to the following constraint on $\alpha$ and $\beta$:
\beq
\Hor^+: \int d\theta^A\sqrt{|\tilde{q}|} (N_\phi + \beta + \eta (\alpha + \beta)) = 0\ \ \   \rightarrow\ \ \  \beta + \eta(\alpha + \beta) =- \frac{8 \pi G J_H}{ A},
\label{tl}
\eeq
in which $ \sqrt{|\tilde{q}|}=  \sqrt {|q_{AB}|}|\psi|$, and we have used the results of the field variations  $\delta_\zeta \ell^a, \delta_\zeta \kappa$ in (\ref{dl}), (\ref{dkappa})  together with the expressions for area and angular momentum (\ref{J}) to arrive at the above result. At first sight, one might worry whether such boundary condition will eliminate the central extension as well. However, notice that
\beq
\mathbf{O}_\eta (g,\pounds_{\zeta_{-m}} g_{ab}) \propto e^{- i m \phi},\ \ f_m \propto e^{i m \phi}, 
\eeq
hence the \textit{only} obstruction term that such boundary condition will not annihilate is of the following form:
\beq
\iota_{\zeta_{n}} \mathbf{O}_{\eta} (g, \pounds_{\zeta_{m} } g_{ab})  \delta_{m+n,0}\neq 0,
\label{left} 
\eeq
as $\delta_{m+n,0}$ made sure that such term does not contain the phase factor $e^{i m \phi}$, which would have made the integration of an axisymmetric function  vanish on $\Sph$. From (\ref{K}), it is immediate to see that the commutator of (\ref{left}) precisely gives the central extension   after imposing (\ref{tl}):
\beq\label{Kf}
\begin{split}
	\Hor^+:  \mathcal{K}_{m,-m} 
	&=	\frac{1}{16 \pi G}\int_{\Sph}  \left(\mathbf{O}_\eta (\pounds_{\zeta_{-m}} g) f_m-  \mathbf{O}_\eta (\pounds_{\zeta_{m}} g) f_{-m} \right) \epsilon_\Sph \\
	&= i m^3 (\eta +1) \frac{ A}{4 \pi G (\alpha+\beta)} .
\end{split}
\eeq
Hence the central charge for the Virasoro on $\Hor^+$ is proportional to the area $A$ and also depends on the coefficient of the boundary Lagrangian:
\beq
c_L =(\eta+1)  \frac{3   A  }{  (\alpha + \beta) \pi G}.
\eeq
Note that although the result  has the same coefficient $(\eta +1)$ as the piece of field space exact form $ (\eta +1) \delta (\kappa \epsilon_\Hor)$ in the symplectic potential, all the terms in (\ref{obst}) contribute to $\mathcal{K}_{m,-m}$ and (\ref{tl}) relates them in a specific way to combine into (\ref{Kf}).   This is different with \cite{Chandrasekaran:2020wwn}, in which the central extension purely localized on the noncovariance of the boundary term $\bm{l}_b$.  Interestingly, there is also no linear term on $m$  in $\mathcal{K}_{m,-m}$ after  (\ref{tl}) is imposed. 

An equivalent evaluation can be carried out for the past horizon. The integrability condition  gives us another constraint between $\alpha$, $\beta$, spin $J_H$ and area $A$: 
\beq
\Hor^-:  \int d\theta^A \sqrt{|\tilde{q}|} (- N_\phi + \alpha + \eta (\alpha + \beta)) = 0 \ \ \ \rightarrow\ \ \  \alpha + \eta (\alpha + \beta) = \frac{8 \pi G J_H}{ A}.
\label{tr}
\eeq
The central extension is evaluated under the above condition:
\beq
\Hor^-:	\mathcal{K}_{m,-m} 
	=   i m^3 (\eta +1) \frac{ A}{4 \pi G (\alpha+\beta)} ,
\eeq
which is exactly the same expression as eq.(\ref{Kf}). Hence the two central charges are equal when the Hamiltonians are integrable on each horizon respectively:
\beq
c_L = c_R.
\eeq
 To ensure the conditions (\ref{tl}) and (\ref{tr}) could both be satisfied at the same time, the following constraint must be true:
\beq
\eta = -1/2, \ \ \alpha -  \beta = 2 \pi (T_L - T_R) = \frac{16 \pi G J_H}{ A} .
\label{specialtem}
\eeq
Hence we have derived the special temprature proposed in \cite{Chen:2020nyh} as a necessary condition for integrability without the need of counterterms.  From the geometrical point of view, it selects a particular family of the $AdS_3$ folia by the  relationships on $\alpha$ and $\beta$. The coefficient $\eta = -1/2$ agrees with the result in \cite{Chandrasekaran:2020wwn} with a different derivation.  
The pair of CFT tempratures for BTZ black hole \cite{Maldacena:1998bw} satisfy such condition, while the pair in Kerr/CFT  does not \cite{ Haco:2018ske, Castro:2010fd}.

Under the integrability conditions (\ref{bdry}) (\ref{specialtem}), the two central charges have the following expression and  the Cardy entropy coincides with the Beikenstein-Hawking entropy,
\beq
c_L = c_R = \frac{3 A  }{  2 \pi G (\alpha + \beta) }.
\eeq
Notice that the above expression  could also be expressed in a form that is closely related to the Brown-Henneaux central charge:
\beq\label{cBH}
c_L=c_R =  \frac{3 \int L_{AdS_3} \sqrt {|\tilde{q}|} d\theta^A  }{2G } ,
\eeq
in which $L_{AdS_3} = \frac{2 |\psi|} {\alpha + \beta}$ is the radius  of the $AdS_3$ folia. It enters the expression through
an integration along the transverse dimension ($\theta^A$). 




\subsection{The Virasoro charges}


After properly treating the non-integrable piece (\ref{obst}) to ensure that the Hamiltonians could be well-defined, their density can be evaluated explicitly
\beq
\begin{split}
\bm{H}_{m} &=U^{\frac{im}{\alpha}} e^{i m \phi}  \frac{m^2 - 2 \alpha^2 \eta - i m (\alpha + 4 \alpha \eta + 2 \beta\eta ) }{16 \pi G (\alpha + \beta)} \epsilon_S \\
&\overset{\eta = - \frac{1}{2}} {\longrightarrow}  \frac{1}{16\pi G} U^{\frac{im}{\alpha}} e^{i m \phi} \left(im + \frac{m^2 + \alpha^2  }{\alpha + \beta}   \right).
\end{split}
\eeq
 When we integrate the density $\bm{H}_m$ on the cross-section $\Sph$, due to the factor $e^{i m \phi} $,  all higher modes $m\neq 0$  vanish apart from the zero mode:
\beq
H_{0} = \frac{-\eta}{8 \pi G}\int_\Sph \frac{ \alpha^2 }{ \alpha + \beta }  \epsilon_S \overset{\eta = - \frac{1}{2}} {\longrightarrow}    \frac{\ \alpha^2 A }{16 \pi G(\alpha+\beta)}. 
\eeq
For completeness, we also provide the expression for the charge density on the past horizon when $\eta = -1/2$

\beq
	\bar{\bm{H}}_{m} = \frac{1}{16\pi G} V^{-\frac{im}{\beta}} e^{-i m \phi}  \left(- im + \frac{m^2 + \beta^2  }{\alpha + \beta}   \right),
\eeq
with the zero mode
\beq
\bar{H}_{0} =   \frac{\ \beta^2 A }{16 \pi G(\alpha+\beta)} .
\eeq
The vanishing of the integration on $\Sph$ does not impact the quasilocal charge $H_m$ to be a well-defined generator, as we only need its partial derivatives of the canonical variables to study the flow it generates on the phase space. Same as in $AdS_3/CFT_2$, the difference of zero mode generates rotation of the spacetime:
\beq
H_0 - \bar{H}_{0} = J_H.
\eeq

Due to the closed ambiguity $\bm{\theta} \rightarrow \bm{\theta}  - d\bm{b} (g,\delta g)$ as we have discussed in the  section 2, one could also add a corner term to the symplectic potential,

 
 \beq
 \Theta_{\Hor} \rightarrow  \Theta_{\Hor} - \lambda \int_{\Sph} (\delta \tilde{q} - \delta h) \iota_\ell \epsilon_H.
 \eeq
When $\lambda = 1/2$, such an additional corner term will make the full symplectic potential become a bulk integration on the horizon; the divergence of the boundary charge  matches with the sum of both matter and gravitational energy momentum flux  \cite{Hopfmuller:2018fni}.  Such form of an additional corner term will shift the density of  the Virasoro charges $\bm{H}_{m} \rightarrow \bm{H}_{m} - \Delta   \bm{H}_{m}$ by,
\beq
\Delta   \bm{H}_{m} =\lambda  m  \ U^{\frac{im}{\alpha}} e^{i m \phi}  \frac{m - i \alpha}{8 \pi G(\alpha + \beta)}  \epsilon_S.
\eeq
Both of the zero modes  and the central extension are invariant under consideration of such type of the closed ambiguity:
\beq
\Delta   H_0 = \Delta   \bar{H}_0=0,  \ \ \Delta \mathcal{K}_{m,-m} =0.
\eeq

\section{Discussion}
In this paper, by analyzing the symplectic structure on the null hypersurface, we derived a weak boundary condition  (\ref{bdry}) to ensure the Virasoro charges are integrable Hamiltonians on the covariant phase space. Such condition  gives rise to a special relationship between the two temperatures of the putative CFT and selects a particular family of $AdS_3$ folia in the near horizon region. It leads to two equal central charges  $c_L = c_R$ on the past and future horizon respectively. 

We want to hightlight three important issues which demand a deeper understanding in the future work:  Firstly, there is only one copy of Virasoro algebra defined on each horizon. The Hamiltonians could be well-defined on any spatial cross-section but not on the bifurcation surface $\Bif$.  Although the value of two central charges have a proper limit simultaneously on $\Bif$,  strictly speaking, the Virasoro algebra are not defined on $\Bif$ but the Cardy formula is applied there and gives a correct entropy counting.   

Secondly, to ensure the integrability of charges on both horizons, one needs to choose a specific coefficient $\eta$ for the Lagrangian boundary term $\bm{l}_{b}$  which corresponds to a mixture of  polarization in the spin-0 degree of freedom (see section \ref{sectionkilling}). It will be nice to figure out other reasons regarding such choice from a different aspect of consistency.

 Thirdly, the  integrability condition only specifies the CFT tempratures by the linear relationship (\ref{specialtem}). There is still a freedom on choosing $\alpha$ and $\beta$, while in the literature of  analyzing scattering of a test field on the Kerr background, it was conjectured that the pair of CFT tempratures for Kerr/CFT should be unique \cite{Castro:2010fd, Compere:2012jk, Castro:2013kea, Castro:2013lba}.   The main difference here is that our analysis focuses on the near horizon spacetime rather than the global strucutre of a solution. The inner horizon  of Kerr, as well as the asymptotic far region of the spacetime are substantial on constraining the unique pair of tempratures in the methods of \cite{Castro:2010fd, Castro:2013kea, Castro:2013lba}, but they are not in the regime of the near horizon analysis.   Such freedom allows us to apply the Cardy formula within the range of its validity $T_{L,R} \gg 1$, which has been a main issue in literature. Eventually, the nonuniqueness of $\alpha$ and $\beta$ should be justified by the matching of scattering data of  probe fields with correlation functions in CFT, which will be analyzed in  \cite{toappear}. Also, it will be of great physical interest to study what kind of physical process would implant Virasoro hairs on the horizon.


In this approach, the most curious mystery lies in the fact that Cardy formula for 2d CFT could remarkably capture the horizon entropy in any $d\geq 3$ dimensions. The first hint it teaches us is the possibility of dimensional reduction of the  near horizon quantum gravity.  For Schwarzchild and deSitter horizons, as well as Perry-Myers black hole, we have the freedom of choosing any axial killing vector for the construction and the resulting Virasoro algebras will be isomorphic to each other.  There is no prefered axis. The physical relevant quantities such as the boundary condition (\ref{bdry}) and the central charge (\ref{cBH}) are all expressed as an integration along the transverse dimension relative to the $AdS_3$ folia.   
The second hint this approach teaches us  is the substantial role of modular invariance for the near horizon quantum gravity. The existence of such symmetry for the boundary CFT is the necessary condition to apply Cardy formula in the first place. It will be of great interest to clarify whether modular invariance exists for the gravitational partition function of the near horizon region. We plan to address this question in the future work.  Finally,  this approach hints at a general  ``near horizon/CFT" duality.  For example, for BTZ black hole, if the putative CFT for our near horizon analysis could be well-defined, then it would capture a subset of  degrees of freedom compared to the standard holography, as it does not detect the asymptotic far region of the black hole.  It would be ``a holography within a holography", which will be very interesting to study.




\acknowledgments
 I would  like to thank Venkatesa Chandrasekaran on the coordination of  arXiv releasing day  with the first version of the draft. The analysis of Lagrangian boundary term $\bm{l}_b$ was only added in the 2nd version of the present paper. I'm grateful to Wan Zhen Chua, Laurent Freidel, Florian Hopfm{\"u}ller, Shuwei Liu,  Romain Ruzziconi, Antony J. Speranza and Bruno de S. L. Torres for discussion. I am in debt to Perimeter Institute for the nice visiting hospitality during the long epidemic lockdown. This work is supported by the grant ID\# 61466  from the John Templeton Foundation, as part of the “The Quantum Information Structure of Spacetime (QISS)” Project. The opinions expressed in this publication are those of the author(s) and do not necessarily reflect the views of the John Templeton Foundation.

\end{document}